\documentclass[11pt]{report}  

\usepackage{setspace}
\usepackage[margin=1truein]{geometry}
\usepackage{cmbright}
\usepackage[T1]{fontenc}
\raggedright

\usepackage[utf8]{inputenc}
\usepackage{authblk} 
\usepackage{hyperref} 
\usepackage{color,soul} 
\usepackage{textcomp} 
\usepackage{graphicx}
\usepackage{booktabs} 
\usepackage{array}
\usepackage{multirow}

\title{Whole-brain calcium imaging with cellular resolution in freely behaving \textit{C. elegans}}

\author[1,2,3]{Jeffrey P.~Nguyen}
\author[1,3]{Frederick B.~Shipley}
\author[4]{Ashley N.~Linder}
\author[1]{George S.~Plummer}
\author[1,2]{Joshua W.~Shaevitz}
\author[1,*]{Andrew M.~Leifer}

\affil[1]{Lewis-Sigler Institute for Integrative Genomics, Princeton University, Princeton, New Jersey, USA}
\affil[2]{Department of Physics, Princeton University, Princeton, New Jersey, USA}
\affil[3]{(Equal contributions)}
\affil[4]{Princeton Neuroscience Institute, Princeton University, Princeton, New Jersey, USA	}
\affil[*]{Correspondence should be addressed to A.M.L.~(\href{mailto:leifer@princeton.edu}{leifer@princeton.edu} )}

\begin{document}
\maketitle

\begin{abstract}
The ability to acquire large-scale recordings of neuronal activity in awake and unrestrained animals poses a major challenge for studying neural coding of animal behavior.  We present a new instrument capable of recording intracellular calcium transients from every neuron in the head of a freely behaving \textit{C. elegans} with cellular resolution while simultaneously recording the animal's position, posture and locomotion. We employ spinning-disk confocal microscopy to capture 3D volumetric fluorescent images of neurons expressing the calcium indicator GCaMP6s at 5 head-volumes per second. Two cameras simultaneously monitor the animal's position and orientation. Custom software tracks the 3D position of the animal's head in real-time and adjusts a motorized stage to keep it within the field of view as the animal roams freely.  We observe calcium transients from 78 neurons and correlate this activity with the animal's behavior. Across worms, multiple neurons show significant correlations with modes of behavior corresponding to forward, backward, and turning locomotion. By comparing the 3D positions of these neurons with a known atlas, our results are consistent with previous single-neuron studies and demonstrate the existence of new candidate neurons for behavioral circuits.
\end{abstract}

\section{Introduction}
How does a nervous system control an animal's behavior? To answer this question, it is important to develop new methods for recording from large populations of neurons in animals as they move and behave freely.  The collective activity of many individual neurons appears to be critical for encoding behaviors including arm reach in primates \cite{Maynard1999}, song production in zebrafinch \cite{Long2010}, the  choice between swimming or crawling in leech \cite{Briggman2005}, and decision-making in mice during navigation \cite{Harvey2012}. New methods for recording from larger populations of neurons in unrestrained animals are needed to better understand neural coding of these behaviors and to better understand neural control of behavior more generally. 
 
Calcium imaging has emerged as a promising technique for recording dynamics from populations of neurons. Calcium-sensitive fluorescent dyes or proteins are used to visualize changes in intracellular calcium levels in neurons \textit{in vivo} which serve as a proxy for neural activity \cite{Chen2013}. To resolve the often weak fluorescent signal of an individual neuron in a dense forest of other labeled cells requires a high magnification objective with a large numerical aperture that, consequently, can image only a small field of view. To avoid artifacts induced by animal motion, calcium imaging has traditionally been performed on animals that are stationary from anesthetization or immobilization. As a result, calcium imaging studies have historically focused on small brain regions in immobile animals that exhibit little or no behavior \cite{Kwan2008}.

No previous neurophysiological study has attained whole brain imaging with cellular resolution in a freely behaving animal. Previous cellular resolution calcium imaging studies of populations of neurons that involve behavior investigate either fictive locomotion \cite{Ahrens2012, Briggman2005}, or behaviors that can be performed in restrained animals, such as eye movements \cite{Kubo2014} or navigation in a virtual environment \cite{Dombeck2007}. One exception has been the development of fluorescence endoscopy which allows recording from rodents during unrestrained behavior, although imaging is restricted to even smaller sub-brain regions \cite{Flusberg2008}.  

Investigating neural activity in small transparent organisms allows one to move beyond studying sub-brain regions to record dynamics from entire brains with cellular resolution. Whole-brain imaging was performed first in larval zebrafish using two-photon microscopy, where the activity of individual neurons was inferred through statistical analysis of pixel intensities  \cite{Ahrens2012}. More recently, whole-brain imaging was performed with cellular segmentation in \textit{C. elegans} using two-photon \cite{Schrodel2013} and light-field microscopy \cite{Prevedel2014}.  Animals in these studies were immobilized, anesthetized or both. 

\textit{C. elegans'} compact nervous system of only 302 neurons and small size of only 1mm make it ideally suited for the development of new whole brain imaging techniques. Calcium imaging in \textit{C. elegans} has already been performed in freely moving animals, although only in very small subsets of neurons and without any $z$-sectioning \cite{Clark2006,BenArous2010,Faumont2011,Piggott2011,Kawano2011,Larsch2013,Shipley2014}. These studies, combined with laser-ablation experiments, have identified a number of neurons that correlate or affect changes in particular behaviors including the AVB neuron pair and VB-type motor neurons for forward locomotion; the AVA, AIB and AVE neuron pairs and VA-type motor neurons for backward locomotion; and the RIV, RIB, and SMD neurons and the DD-type motor neurons for turning behaviors \cite{Kawano2011,Faumont2011,BenArous2010,Piggott2011,Faumont2011,Shipley2014,Donnelly2013,Gray2005}. To move beyond these largely single-cell studies, we sought to record simultaneously from the entire brain of \textit{C. elegans} with cellular resolution and record its behavior as it moved about unrestrained.

\section{Results}
\subsection{Volumetric imaging in a freely moving worm}
\begin{figure}
\begin{center}
\includegraphics[width=2.5in]{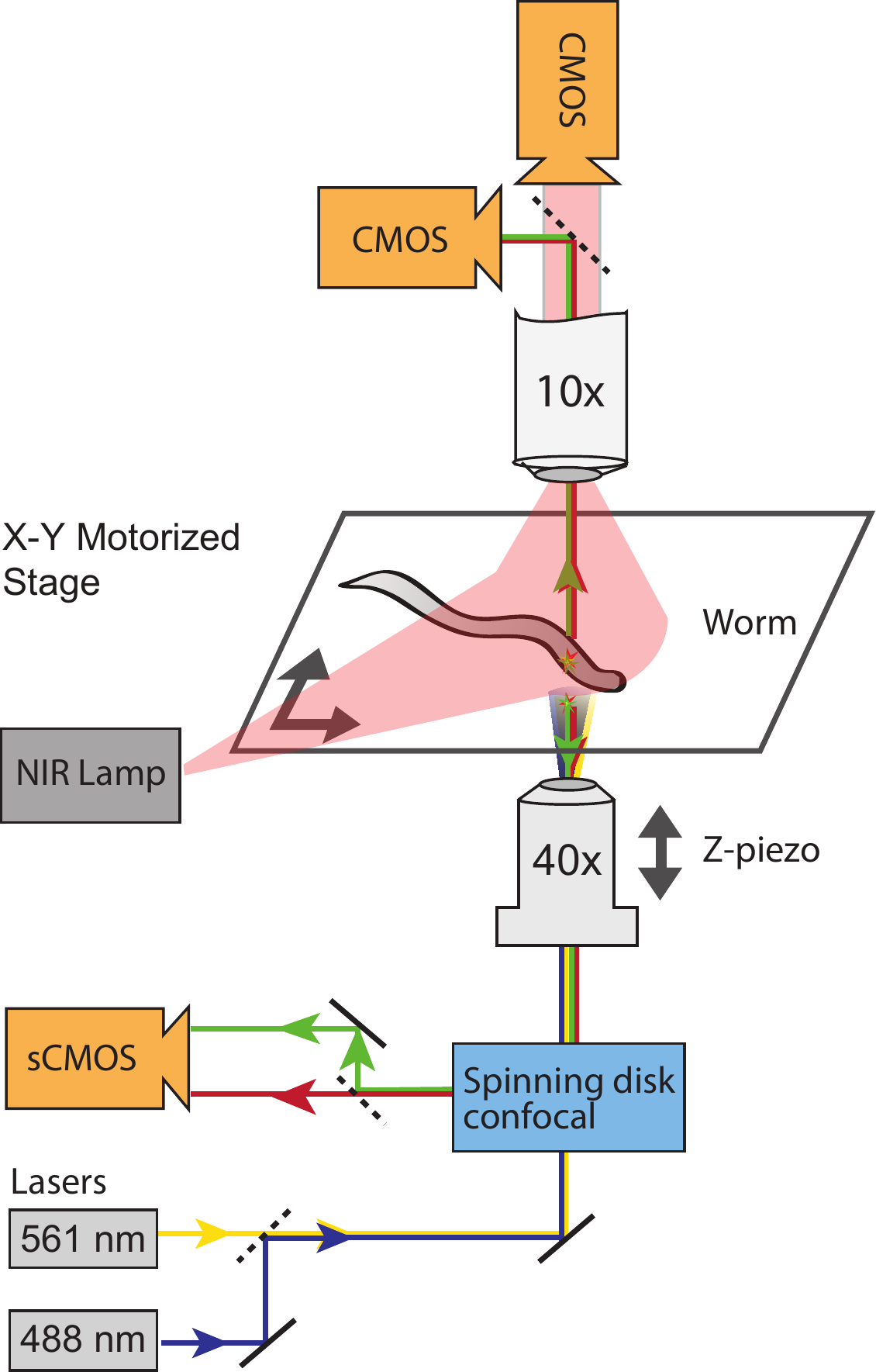}
\end{center}
\textbf{\refstepcounter{figure}\label{fig:optics} Figure \arabic{figure}.}{ | Simultaneous imaging of whole-brain calcium activity and behavior, simplified schematic. A worm crawls freely on a motorized stage under near-infrared  (NIR) dark-field illumination. A spinning disk confocal microscope acquires volumetric fluorescent images of the worm's brain by scanning a 40x objective along the imaging axis ($z$) to acquire 5 brain volumes per second. A low magnification 10x objective images the animal's posture and behavior. Real-time computer vision software tracks the worm in three dimensions and adjusts the $xy$ motorized stage and $z$-piezo to keep the worm's head centered in the high magnification objective's field of view.}
\end{figure}

To record fluorescent images of all neurons in the worm's head while simultaneously recording the animal's behavior, we developed a custom dual-objective tracking instrument. A spinning disk confocal microscope records images of the head of a worm through a 40x objective using both 488 nm and 561 nm excitation laser light while a second imaging path records the worm's position and behavior through a 10x objective (Fig.~\ref{fig:optics}).  Custom real-time computer vision software tracks the worm in three dimensions and adjusts a motorized stage and piezoelectric objective mount to follow the worm and to scan through its brain volume as it crawls.

\begin{figure}
\begin{center}
\includegraphics[width=3.3in]{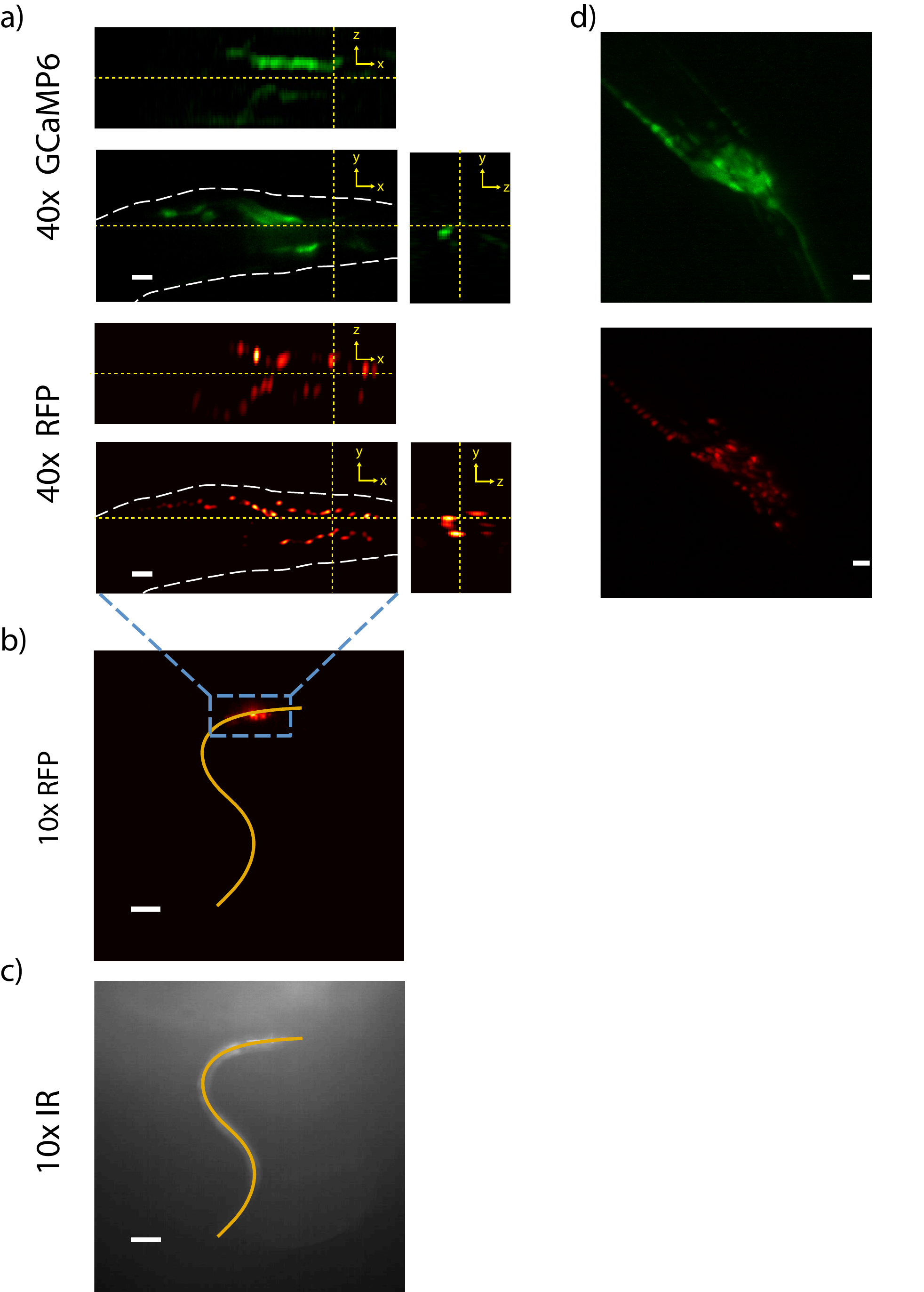}
\end{center}
\textbf{\refstepcounter{figure}\label{fig:fluorFrames} Figure \arabic{figure}. }{| Simultaneous recording of four video streams showing neural activity and behavior. \textbf{a)} GCaMP6s and RFP fluorescence are each recorded at 200 fps through the 40x objective as it scan's through the worm's head in $z$. A 3D volume is reconstructed from a $z$-stack of acquired images. Individual $xy$, $xz$ and $zy$ slices are shown. Slice location is indicated by a yellow dashed line.  An approximate outline of the worm's head is indicated by a white dashed line. Scale bar is 10 $\mu$m. Each $xy$ slice was recorded for GCaMP6s and RFP simultaneously. \textbf{b)} RFP fluorescence is also recorded through a higher depth-of-field 10x objective at 61 fps.  \textbf{c)} The worm's posture and behavior are  recorded via infrared darkfield imaging through the 10x objective at 47 fps.  \textbf{b-c)} Scale bar is 100 $\mu$m. Orange line indicates the worm's centerline. \textbf{d)} Maximum intensity projection of the same fluorescent images shown in \textbf{a,b}. Scale bar is 100 $\mu$m.}  
\end{figure}

The worm expresses a genetically encoded calcium indicator, GCaMP6s, in all of its neurons \cite{Chen2013} and a calcium-insensitive fluorescent protein, RFP, in the nucleus of each neuron. To acquire 3D image stacks of the worm's head, a piezoelectric stage translates the 40x objective up and down along the imaging axis such that the focal plane passes through the animal's head five times per second. The confocal spinning disk rejects out-of plane light and provides optical sectioning. A high-speed, high-sensitivity Scientific CMOS (sCMOS) camera records fluorescent images at 200 fps, thereby capturing five head volumes per second, each consisting of 40 $z$-slices  approximately 1.2~\textmu m apart. To capture both the calcium dynamics and information about the location of cell bodies,  both red- and green-channel fluorescent images are recorded side-by-side on the sCMOS sensor simultaneously. Each $z$-stack spans a  $150\times150 \times 47$~\textmu m volume and is sufficiently large to image all neuronal cell bodies in the head of the animal, including those located from the tip of the nose, through the nerve ring, to  the ventral nerve cord immediately posterior of the pharyngeal terminal bulb (Fig.~\ref{fig:fluorFrames}d).

An animal's head moves dramatically during crawling which poses challenges for volumetric imaging. Worms crawl with a center of mass velocity of up to 0.6~mm/ s and the head can swing side-to-side by as much as 70 \textmu m, corresponding to two head diameters, in one second. To keep the worm's head centered in the field of view of the high magnification objective, we used a second, low-magnification, microscope located on top of the worm to track its movements in real-time and to adjust a motorized stage to compensate for the animal's motion. A high-speed CMOS camera records large depth-of-field fluorescence images of the neurons in the worm's head through a 10x objective at 61 fps (Fig.~2c). A modified version of the CoLBeRT software \cite{Leifer2011}, written in C, analyzes these low-magnification fluorescent images in real-time to track the worm's $x$-$y$ head position  and to adjust the velocity of the motorized stage so as to keep the head of the worm centered. This $x$-$y$ feedback loop updates at 61 Hz and keeps the worm within the field of view for the duration of our recordings. 
	
The worm also moves its head along the imaging axis ($z$) as it crawls. This occurs when the worm actively raises its head off of the agarose pad or when the worm crawls over agarose of varying thickness. To keep the worm's head in the center of the $z$-scanning range, we analyze fluorescent images from the high magnification objective in real-time to find the center of the worm's head in each $z$-stack. We then adjust the piezo scan range to compensate for any head motion. This $z$ feedback loop, written in LabView, updates at 5~Hz and keeps the worm within the scan range. For the data shown here, the feedback loop compensated for changes in the worm's head position of as large as 8~\textmu m in $z$, or 20\% of its head diameter, in the course of a single 20~ms volume acquisition. 

To study neural coding of behavior, it is critically important to record the animal's full body posture. A third camera in our system records the worm's posture and behavior at 47 fps through the 10x objective using near-infrared (NIR) dark-field illumination. Images from this camera are analyzed off-line to define the body shape over time. A sample frame recorded simultaneously from each of the four video feeds is visible in (Fig.~\ref{fig:fluorFrames}).

\subsection{Whole-brain neural dynamics with cellular resolution during locomotion}
\begin{figure}
\begin{center}
\includegraphics[width=4in]{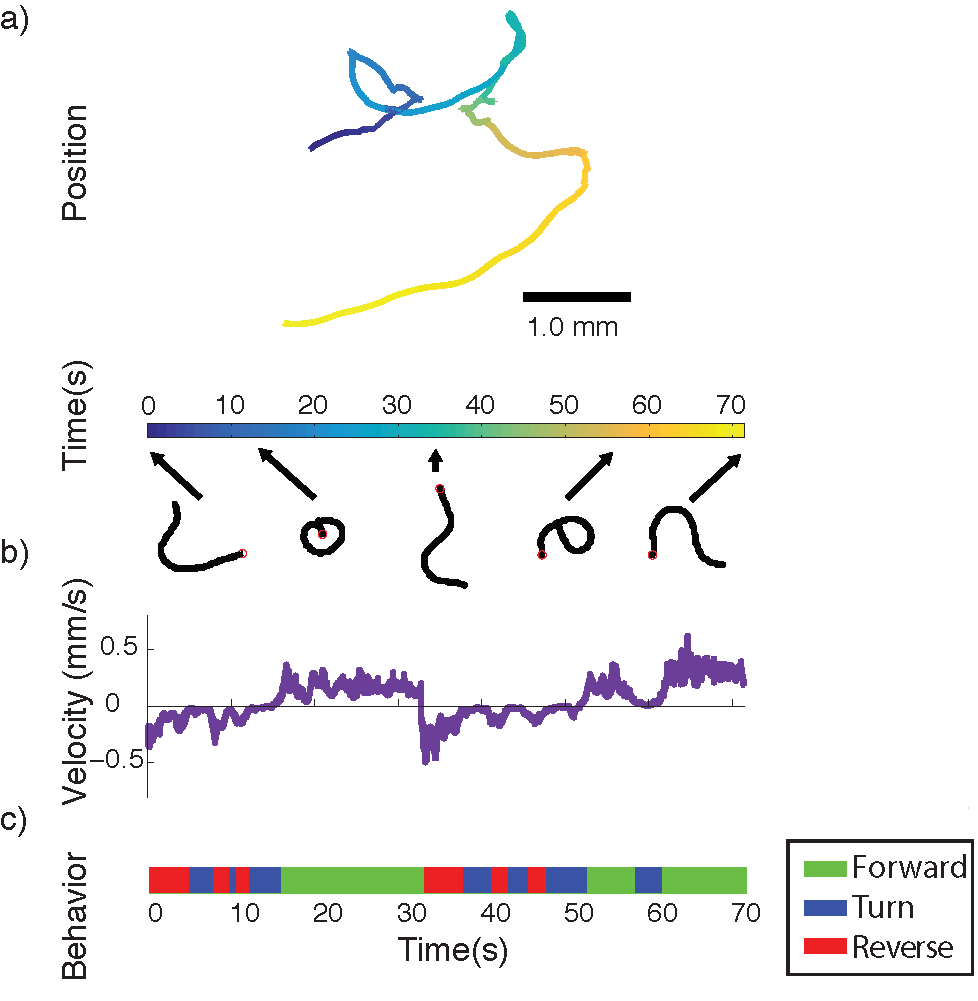}
\end{center}
\textbf{\refstepcounter{figure}\label{fig:behavior} Figure \arabic{figure}. }{| The worm's behavior is recorded during imaging.  Data from Worm 2 is shown. The worm's \textbf{a)}  center-of-mass trajectory, body shape, and  \textbf{b)} velocity in the anterior direction are shown. \textbf{c)} An ethogram describing the behavior is generated automatically from the worm's posture and behavior (see methods). }
\end{figure}

As a proof of principle, we recorded the neural activity and behavior of two individual worms crawling freely on agarose pads. We quantified their behavior by calculating the worm's center-of-mass velocity and body centerline at each time point. This data was then used to define the worm velocity relative to the body orientation such that positive velocity corresponds to forward locomotion and negative velocity to backward locomotion (Fig.~\ref{fig:behavior}). We used the worm's velocity and body posture to extract a discrete set of behaviors to which we later correlated neural activity. Examination of the centerline dynamics established that, for these animals, when the magnitude of the measured center-of-mass velocity was small, the worm executed a series of sharp, body-bending turns. Combining the information from the velocity and the centerline dynamics, we categorized behavior into three modes: forward crawling, backward crawling, and turning.

To extract calcium dynamics from our 3D image stacks, we measured the fluorescence intensity of GCaMP6s in each neuron in a region defined by the RFP fluorescence in that neuron's nucleus. We chose to express RFP only in the cell's nuclei to help distinguish each cell from its neighbor. Nuclear regions were identified through a combination of manual selection and computer vision analysis. This procedure resulted in the identification of 78 and 68 neuronal nuclei in the two worms, accounting for 45\% and 50\% of the 181 neurons expected to be in tour field of view \cite{White1986,wormbase}. These numbers are comparable to the $83\pm 15$ (mean plus or minus standard deviation) neurons identified in a recent whole-brain imaging study of anesthetized worms \cite{Schrodel2013}.

\begin{figure}
\begin{center}
\includegraphics[width=\textwidth]{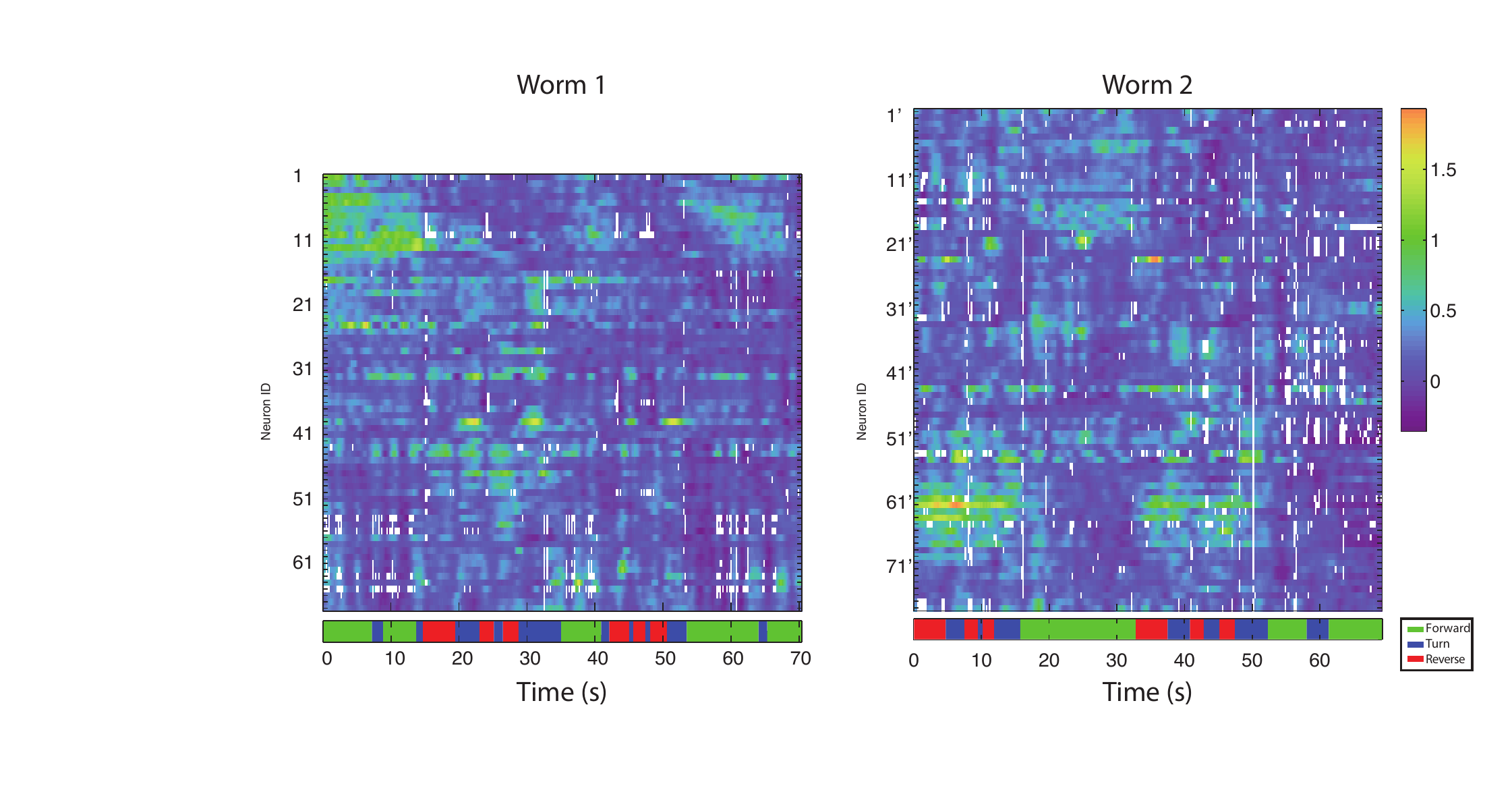}
\end{center}
\textbf{\refstepcounter{figure}\label{fig:heatmap} Figure \arabic{figure}. }{| Neural activity from 78 and 68 neurons from two individuals were recorded during behavior. GCaMP6s fluorescence from all  neurons is plotted. Fluorescent intensity is represented as the fractional change in fluorescence $\Delta F/F_0$ where the baseline $F_0$ is defined for each neuron as the lower 20th percentile value. Neurons are sorted via a hierarchical clustering algorithm. Corresponding correlations are shown in Supplementary Figure  S\ref{fig:corrMatrix}. Behavioral ethograms are shown. Neural data from Worm 2 corresponds to the same recording as in Figure \ref{fig:behavior}.}
\end{figure}

Each neuron's GCaMP6s fluorescence was measured in a segmented region within an 8x8x8~\textmu m$^3$ volume centered around the nucleus. The segmented region itself was defined by fluorescence from the RFP channel. Fluorescence intensity of each neuron was sampled at 5 Hz, similar to the 4-6 Hz used in previous whole-brain imaging experiments on immobilized and anesthetized worms (Figs.~\ref{fig:heatmap},~S\ref{fig:exampleTraces},~S\ref{fig:corrMatrix}) \cite{Schrodel2013}. This rate is sufficient to capture a vast majority of the calcium dynamics probed with GCaMP6s  (Fig. S\ref{fig:PSD}). The most active neurons exhibited calcium transients with a $\Delta F /F_0$ of  150\%, where $F$ is the intensity of the of GCaMP6s fluorescence and $F_0$ is the neuron-specific fluorescence baseline (see Methods). We were able to record calcium dynamics even as the worm turned, reversed, touched itself or reoriented. 

\subsection{Specific neurons correlate with each of the three observed behaviors}
%
%

We next calculated the correlation between the dynamics of individual neurons and changes in locomotory behaviors. The $\sim80$ neurons we observed in each worm displayed a varying degree of correlation with each of the three behaviors we measured. To reveal neurons with significant behavioral correlations, we compared results with a shuffled noise model and used baseline estimation (see Methods). This yielded at least two neurons for each worm that were correlated with each of the three behaviors with an average coefficient of $r=0.50$.

\begin{figure}
\begin{center}
\includegraphics[width=6in]{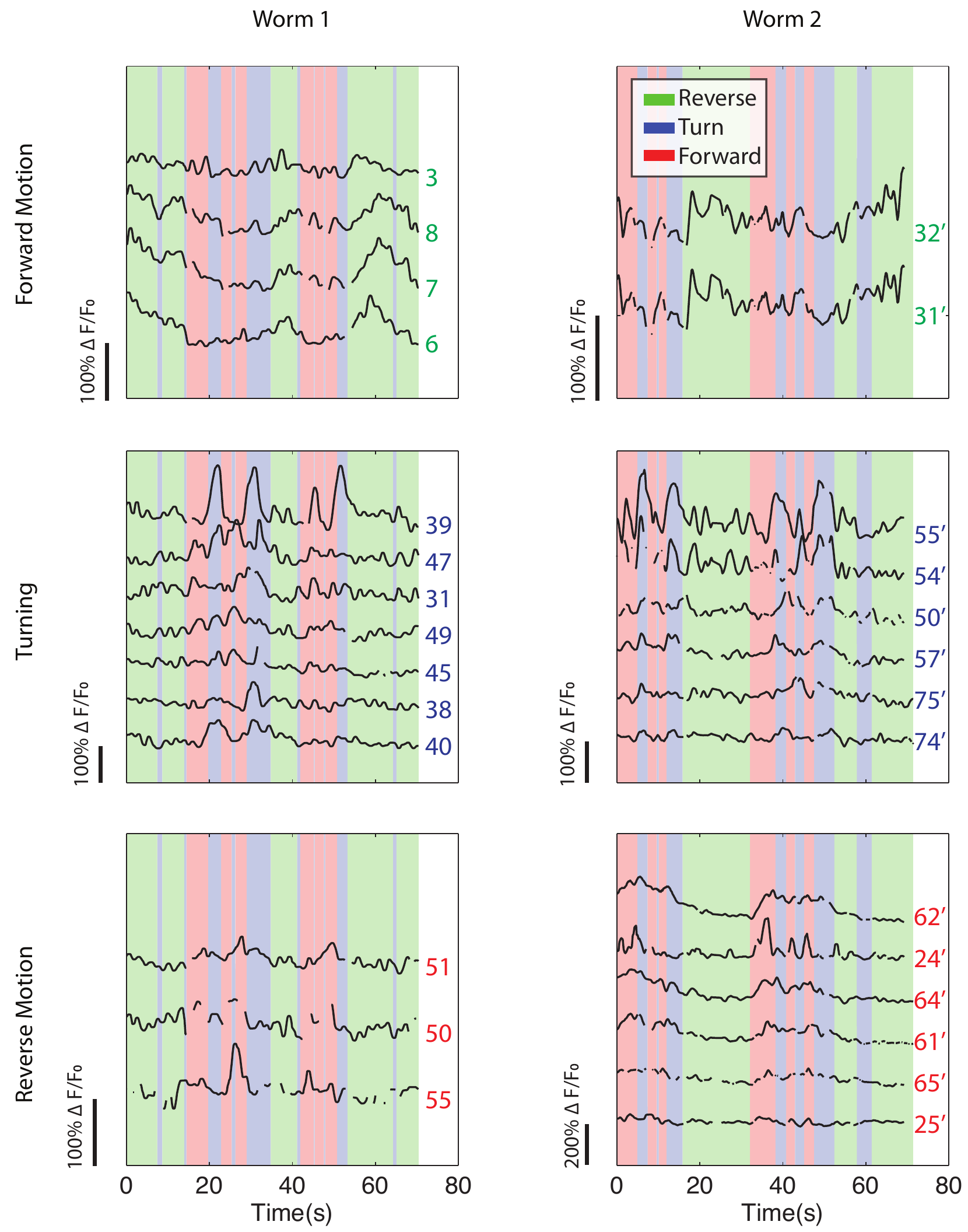}
\end{center}
\textbf{\refstepcounter{figure}\label{fig:behaviortraces} Figure \arabic{figure}. }{| All neurons whose activity correlated with either forward, turning or reverse behavior above a significance threshold are shown. Each neuron's ID number is indicated.}
\end{figure}

We observed three and six neurons in each worm, respectively, whose calcium activity correlated with backward motion. We observed seven and six neurons in each worm, respectively, whose calcium activity correlated with turning locomotion, in which we group so-called ``omega turns,'' ``kinks'' and other large curvature events.  Finally we observed four and two neurons in each worm, respectively, whose calcium activity correlates with forward locomotion. Interestingly, the numbers of neurons we see involved in each behavior are roughly similar to the numbers suggested by other evidence in the literature, see Supplementary Table S\ref{table:NeuronEvidence}. 

One potential concern is that the neural activity we observe could be an artifact of the worm's body motion instead of changes in intracellular calcium levels. First, we note that forward and backward locomotion correspond to the same cyclic motion of the body, only reversed in time. The fact that we find separate sets of neurons whose activity correlates with these behaviors indicates that the dynamics are likely not due to motion artifact, regardless of the neuron's position. Second, we used the nuclear-localized RFP fluorescence as a built-in control for motion artifacts as it is insensitive to the levels of intracellular calcium. RFP fluorescence dynamics from neurons whose GCaMP6s signal most correlated with the three behaviors did not correlate with behavior, yielding an average correlation coefficient of $r=0.13$ (Supplementary Figure S\ref{fig:rfpcontrol1}, S\ref{fig:rfpcontrol2} ). Third, we repeated the experiment with a GFP control worm and saw significantly smaller changes in fluorescent intensity over time (Supplementary Figure S\ref{fig:GFPhist}). Lastly,  neurons that are neighbors in space show qualitatively different temporal dynamics which would not be expected if the dynamics arose from motion artifacts (Fig. S\ref{fig:exampleTraces}).

\begin{figure}
\begin{center}
\includegraphics[width=\textwidth]{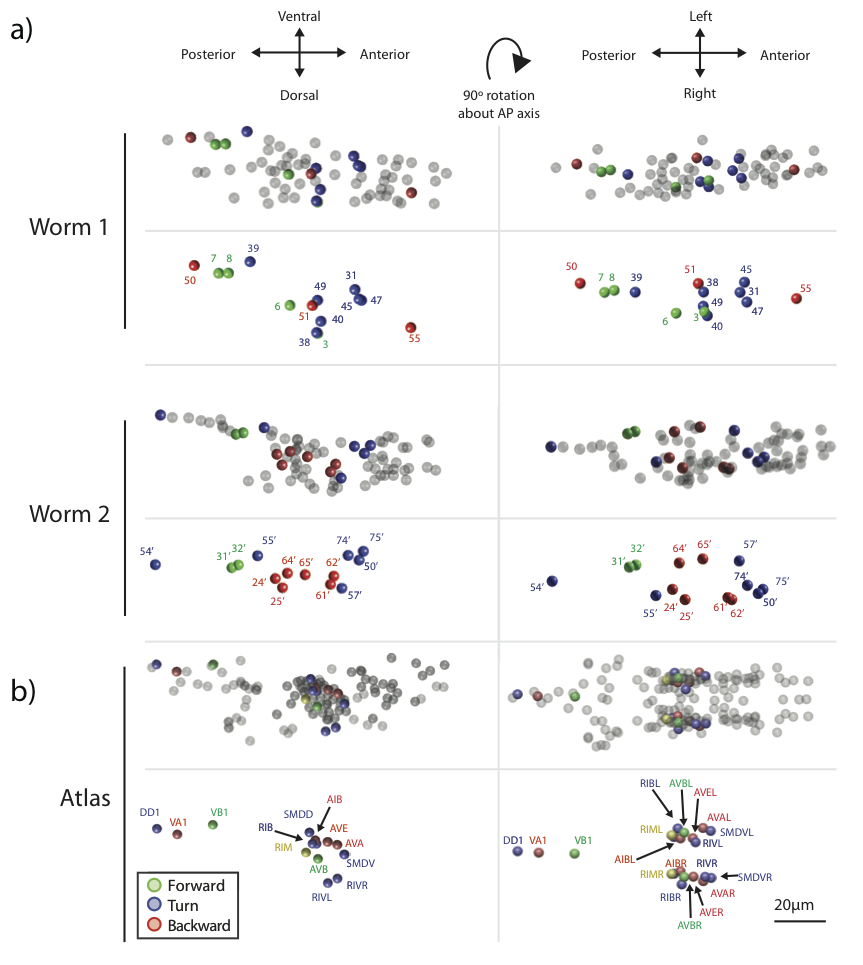}
\end{center}
\textbf{\refstepcounter{figure}\label{fig:3Datlas} Figure \arabic{figure}. }{| The position of recorded neurons for each worm  is plotted in 3D and compared to an atlas. The left column shows a dorsal-ventral plane view of the animal, while the right column shows  left-right plane view of the animal. \textbf{a)} Neurons whose activity correspond to forward, turning or reverse are colored and labeled labeled with a neuron ID corresponding to Figure \ref{fig:heatmap}. These neurons are shown both in and out of context of all other recorded neurons, which are shaded gray. \textbf{b)} The position of all known neurons are plotted using reference data culled from the WormBase Virtual Worm Project. Neurons are colored according to their suspected role in behavior based on evidence in the literature (see Supplementary Table S\ref{table:NeuronEvidence}) and  labeled with their neuron name. }
\end{figure}

\begin{figure}[htb]
\begin{center}
\includegraphics[width=5in]{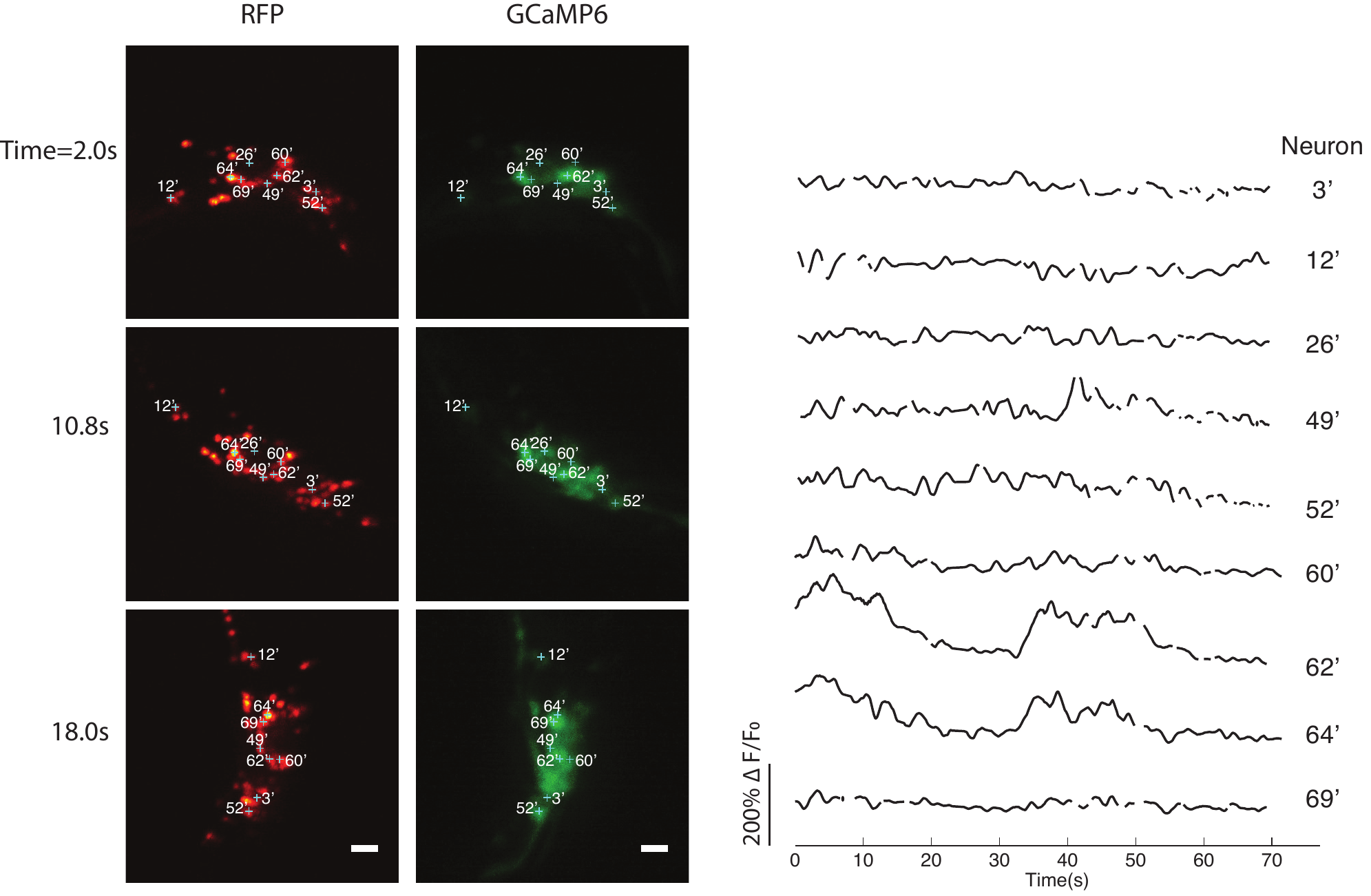}
\end{center}
\textbf{\refstepcounter{Sfig}\label{fig:exampleTraces} Supplementary Figure S\arabic{Sfig}. }{| Example traces recorded during behavior. Calcium transients from 17 neurons located in a horizontal slice of the worm are shown. Each neuron's position is indicated in the RFP and GCaMP6s fluorescent images. Data is from Worm 2 and corresponds to recordings shown in Figures \ref{fig:behavior} and \ref{fig:heatmap}. Fluorescent images show maximum intensity of three $z$-planes.}
\end{figure}

We next sought to use the three-dimensional positions of the nuclei in our images to identify the neurons that correlate with behavior. Plotting the neuronal positions, we see a striking correspondence between the behavior-specific neurons in the two worms we probed (Fig. \ref{fig:3Datlas}). While not an identical overlap, we identify at least seven forward, backward, or turning-specific neurons that are in similar locations in the two worms. 

As in previous whole-brain imaging studies \cite{Schrodel2013,Prevedel2014}, we do not know the true identities of the neurons we observe. We instead identified neurons from the literature that affect the three locomotory behaviors (See Supplementary Table S\ref{table:NeuronEvidence}). We plotted their positions in three-dimensions (Fig.~\ref{fig:3Datlas}) using data culled from the WormBase Virtual Worm Project \cite{wormbase} which in turn is based on electron microscopy data from \cite{White1986}.  Comparing this atlas image to data from our two worms allowed us to assign plausible neural identities to the neurons we observed who had activity correlated with forward, turning and reversing.  

The position of our forward neurons are consistent with having observed the neural dynamics of AVBR (neuron  \#3), and VB1 (neuron  \#7 or 8 and \#31' or 32'). The neural dynamics observed in these neurons match  prior calcium imaging studies that have implicated the AVB neuron pair  and the VB type motor neurons  in backward locomotion, although calcium activity of VB1 itself has never  been observed \cite{Kawano2011,Piggott2011}.  The position of our reversing neurons indicates that we very likely observed the dynamics of VA1 (neuron \#50). Previous calcium imaging studies have implicated the VA type motor neurons in backward locomotion, although VA1 calcium dynamics has never been directly measured \cite{Kawano2011}. We also observe a cluster of reverse neurons whose position is similar to a the group of neuron pairs, AVA, AVE and AIB, that have  previously been observed to have  elevated calcium levels during backward motion \cite{BenArous2010,Kawano2011,Piggott2011,Faumont2011,Shipley2014}. Neurons \#51, 64' and 65' likely correspond to any of AVEL, AVAL and AIBL.  Three of neurons \#24', 25', 61', and 62' likely correspond to AVAR, AVER and AIBR.   We also observed neurons whose activity correlated with turning.  In  the literature, calcium activity in non-sensory neurons in the head has never been observed to correlate with turning, so we looked for non-sensory neuron candidates based on laser ablation studies \cite{Gray2005,Donnelly2013} to identify candidate turning neurons. Neuron \#38 likely corresponds to RIVL,  neuron \#57'  to SMDVL and neuron 54' to DD1.

\section{Discussion}
The imaging system presented here allows for whole-brain recording of calcium activity with cellular resolution in a freely moving \textit{C. elegans}. We used this system to investigate population dynamics of neurons in the worm whose activity correlates with forward, backward, and turning modes of locomotion. Some of these neurons compare favorably with results from the literature whereas others appear to  have new roles in locomotory behavior. This is also the first example of recorded calcium dynamics from non-sensory neurons in the head that correlate with turning, where we identify DD1, SMDV and RIV along with other candidate neurons. We believe this work represents a significant advance towards  studying population dynamics of a brain-sized neural network for coding behavior.

One of the advantages of working in \textit{C. elegans} is that its neural connectivity has been mapped \cite{White1986}. Unfortunately, there is currently no principled way of relating whole-brain, three-dimensional images to the known neuroanatomical atlas. This challenge has plagued imaging studies in adult \textit{C. elegans} in the past \cite{Schrodel2013,Prevedel2014} and remains unresolved here. Future work is needed to solve this neural identity challenge.

We currently record the fluorescence from genetically-encoded calcium indicators using spinning disk microscopy. However, our system is fully modular and the real-time tracking microscopy  and off-line image analysis are  well suited to take advantage of other imaging modalities such as light-field or two-photon microscopy, as well as advances in genetically-encoded voltage indicators \cite{Hochbaum2014,Jin2012,Gong2013}.

\section{Acknowledgements}
This work was supported by a grant from the Simons Foundation (SCGB \#324285) to A.M.L.~and by Princeton University's Inaugural Dean's Innovation Fund for New Ideas in the Natural Sciences to J.W.S.~and A.M.L. A.N.L.~is supported by a National Institutes of Health institutional training grant through the Princeton Neuroscience Institute. We thank M.~Zimmer for productive discussions and M.~Zimmer and O.~Hobert for sharing reagents used in preliminary work for this study.  We thank C.~Clark and M.~Alkema for kindly gifting their GCaMP6s strain, and A.~Sylvain and C.~Murphy for gifting their GFP strain. 


\section{Author contributions}
A.M.L.~conceived the project with input from J.W.S. The authors F.B.S., A.M.L., and J.W.S.~designed and built the hardware setup. J.W.S., J.P.N., A.M.L.~and F.B.S.~wrote the real-time instrument control software. G.S.P.~conducted molecular genetics. F.B.S., J.P.N., A.N.L.~and G.S.P.~conducted experiments. J.P.N.~and A.N.L.~wrote image processing software and J.P.N., F.B.S., A.N.L.~and J.W.S.~performed analysis. A.M.L.~and J.W.S.~wrote the manuscript with input from all authors.

\section{Methods}
\subsection{Strains}
We cultivated transgenic worms in the dark at 20\textdegree C on nematode growth medium (NGM) plates with OP50 bacteria. OP50 plates were made by seeding 6-cm NGM plates with 250 \textmu l of a suspension of OP50 bacteria in LB broth.

Strain QW1217 (zfIs124(P\textit{rgef-1::GCaMP6s}); otIs355(P\textit{rab-3::NLS::tagRFP})) was a gift of C.~Clark and M.~Alkema. 
Strain AML10 (otIs45(P\textit{unc-119::GFP}); otIs355(P\textit{rab-3::NLS::tagRFP}) was generated by crossing a male of strain QW1155 (otIs355(P\textit{rab-3::NLS::tagRFP})) with a hermaphrodite of strain OH441 (otIs45(P\textit{unc-119::GFP})). Strain QW115 was a gift of C. Clark and M.~Alkema and OH441 was git of A.~Sylvain and C.~Murphy.

\subsection{Microscopy}
The microscope is built on the base of a Nikon Eclipse TE2000S inverted microscope. A schematic with details of optical elements is shown in Supplementary Fig.~S\ref{fig:fullOptics}. 

Illumination is delivered from below the sample. Dark-field illumination is provided by a halogen lamp filtered with a near-infrared longpass filter (Omega Optical \#3RD710LP) and directed by light guides so as to land obliquely on the sample from below. GCaMP6s and RFP in the sample are excited by 561nm and 488nm light from two diode-pumped solid state lasers (Coherent Sapphire 561 LP and 488 LP, each with 200-mW maximum power). Light from the two lasers are combined using a dichroic mirror (Semrock FF518-Di01-25x36) and then coupled into an optical fiber so as to illuminate the sample through a spinning disk confocal unit (Yokogawa CSU-X1) and 40x objective (Nikon CFI Super Fluor, 1.3 NA). Power at the sample for the GCaMP6s experiments was measured to be 63 mW / mm$^2$ and 48 mW / mm$^2$ for 561 nm and 488 nm light respectively. For the GFP control worm experiment, the blue laser intensity was adjusted such that the recorded GFP fluorescence intensity matched that observed in a GCaMP6s worm.

Fluorescent images of GCaMP6s and the RFP are recorded at high magnification  with a narrow depth-of-field through the spinning disk and 40x objective onto a `Scientific' CMOS (sCMOS) camera (Hamamatsu Orca Flash v 4.0). Three dichroic mirrors ensure that only emitted light of the appropriate frequency make it to the camera, while excitation light is rejected. The first dichroic mirror (Edmund Optics, 650nm SP Dichroic) rejects stray near-infrared light that comes from the dark-field illumination source. The second dichroic (Semrock Di01-T405/488/568/647-13x15x0.5), located in the spinning disk, allows the emitted green and red light to pass, but rejects the excitation laser light. The third dichroic (Semrock FF670-SDi01-25x36), located in a dual imager (Photometrics DV2 DualView), splits the emitted light into red and green channels that are imaged side-by-side on the camera. Additional emission filters (Semrock FF01-520/32 and FF01-609/54) provide further light rejection.

Fluorescent images are also recorded at low magnification and high depth-of-field through the 10x objective above the sample (Nikon Plan Apo 10x 0.45 NA) to facilitate real-time $xy$ tracking. In that imaging path, fluorescent light passes from the objective, through a longpass filter (Semrock LP02-568RU-25) to reject excitation light, and then is reflected by a dichroic mirror (Nikon 660nm LP) and imaged onto a CMOS camera (Basler acA2000-340km). To record animal behavior, darkfield near-infrared light is imaged through the same 10x objective. The near-infrared light  passes through the dichroic to form an image on a second CMOS camera (Basler acA2000-340km-NIR).

\subsection{Image acquisition}
Three  cameras were used to acquire high- and low-magnification fluorescence and behavior images. Their properties are listed in Table \ref{table:cam}. To synchronize the recorded video streams during post-processing, a bright white flash was generated from a consumer camera flash (Bower SFD328) that was simultaneously visible on all three cameras.

Raw uncompressed high-magnification fluorescent images were recorded via custom Labview software to a high bandwidth solid state drive (OCZ Revodrive 350, 1TB), so as to handle the large datastream of 288 MB/s (200fps of 1200x600 pixel 16 bit-depth images). Low magnification fluorescence and behavior images were stored by the \textit{CoLBeRT} software using JPEG lossy compression on standard hard drives \cite{Leifer2011}.

\begin{table}[htbp]
  \caption{Properties and parameters for the three cameras used. \label{table:cam}}
    \begin{tabular}{|l||c|c|c|}
    \hline
    Camera Use & High Mag Fluorescence & Low Mag Fluorescence & Low Mag Behavior \\
    \hline
    Brand & Hamamatsu  & \multicolumn{2}{c|}{Basler} \\
	 \hline
    Model & Orca Flash 4.0  & acA2000-340km  & acA2000-340km-NIR \\
	 \hline
    Sensor & sCMOS & \multicolumn{2}{c|}{CMOS} \\
	 \hline
    Interface & \multicolumn{3}{c|}{Camera Link} \\
	 \hline
    Frame Grabber & BitFlow Karbon PCIe x16 10-tap   & \multicolumn{2}{c|}{Active Silicon Firebird PCIe GEN II 8x} \\
	 \hline
    Active Pixels & 1200 x 600  & 1024 x 768 & 1088 x 1088 \\
	 \hline
    Bit Depth & 16  & \multicolumn{2}{c|}{10} \\
	 \hline
    Image Scale & 4000 px/mm & 580 px/mm & 580 px/mm \\	
	 \hline
    Exposure (ms) & 5     & 10    & 4 \\
	 \hline
    Frame Rate (fps) & 200   & 47   & 61 \\
    \hline
    \end{tabular}%
  \label{tab:addlabel}%
\end{table}%


\subsection{Volumetric imaging and 3D tracking}
The 40x objective is mounted on a piezoelectric stage (Mad City Labs Nano-F200S) that allows it to translate along $z$ under computer control. To acquire volumetric fluorescent images, the objective is driven by a 2.5 Hz triangle wave with a 4 V peak-to-peak voltage from a function generator (Hewlett-Packard HP33120A) so that the objective passes through the sample five times per second. 

The animal crawls in the imaging arena atop an $xy$ motorized stage with linear position encoders (Ludl BioPrecision2). As the worm crawls, a modified version of the \textit{MindControl} software (\url{http://git.io/colbert}) monitors the low-mag fluorescent video stream, locates the worm's head in $xy$ by finding the centroid of the worm's bright fluorescent brain, and then adjusts the velocity of the motorized stage to compensate for the worm's motion \cite{Leifer2011}. This $x$-$y$ feedback loop updates at 47 Hz. 

To compensate for $z$-motion, custom Labview software implements a second feedback loop that monitors the high-magnification fluorescent images and adjusts the objective's $z$-position to track the worm's head. After each pass of the objective (5 Hz), the software estimates the center of the worm's head in $z$ by finding the image plane that has pixel values with the highest standard deviation. The software monitors the $z$-piezo's position through a DAQ board  (National Instruments PCI-6251) and adjusts the $z$-piezo's DC offset after each stack to reflect any head motion by sending an offset voltage via the DAQ that is added to the piezo's driver by a simple analog circuit.

\subsection{Free behaving experiments}
The imaging arena consists of a 100 mm diameter petri dish, turned upside with an NGM pad \cite{Stiernagle2006} where agarose has been used instead of agar upon which the worm crawled. A 50 x 45 mm wide cover slip covered the worm and rested on 127 \textmu m spacers, and mineral oil filled the area between the agarose pad and the cover slip, so as to help with index matching \cite{Leifer2011}. Young adults were picked, washed in M9 buffer, and were placed on the imaging arena and allowed to acclimate for 5 minutes before imaging. The worm crawled freely in the arena during imaging. 

\subsection{Immobilized experiments}
In immobilized experiments, the worm was immobilized on an agarose pad using polystyrene beads \cite{Kim2013}. Basal neural activity was recorded as in the freely-moving experiments, however, recording was performed at 200 fps and  only a single image plane was recorded.

\subsection{Extracting neural signals}

Cell bodies were located by inspecting the nuclear-localized RFP fluorescence images. Cells were detected and tracked using a semi-automated approach. First, the coordinates of six neurons were human-annotated and tracked in all time points. Subsequent neuron positions were detected by interpolation with human-supervision. For each frame, an interpolant is created between it and the previous frame using matching annotated neurons. The interpolant is then used to estimate the position of a neuron in the new frame based on its position in previous frames. The position is refined by searching for local maxima in RFP intensities within 10 $\mu m$ of the initial estimate. No point can be assigned to more than a single neuron and each frame is checked manually to ensure predicted positions match the correct neuron. The process is repeated until no more neurons can be consistently detected. 

\begin{figure}[htb]
\begin{center}
\includegraphics[width=5in]{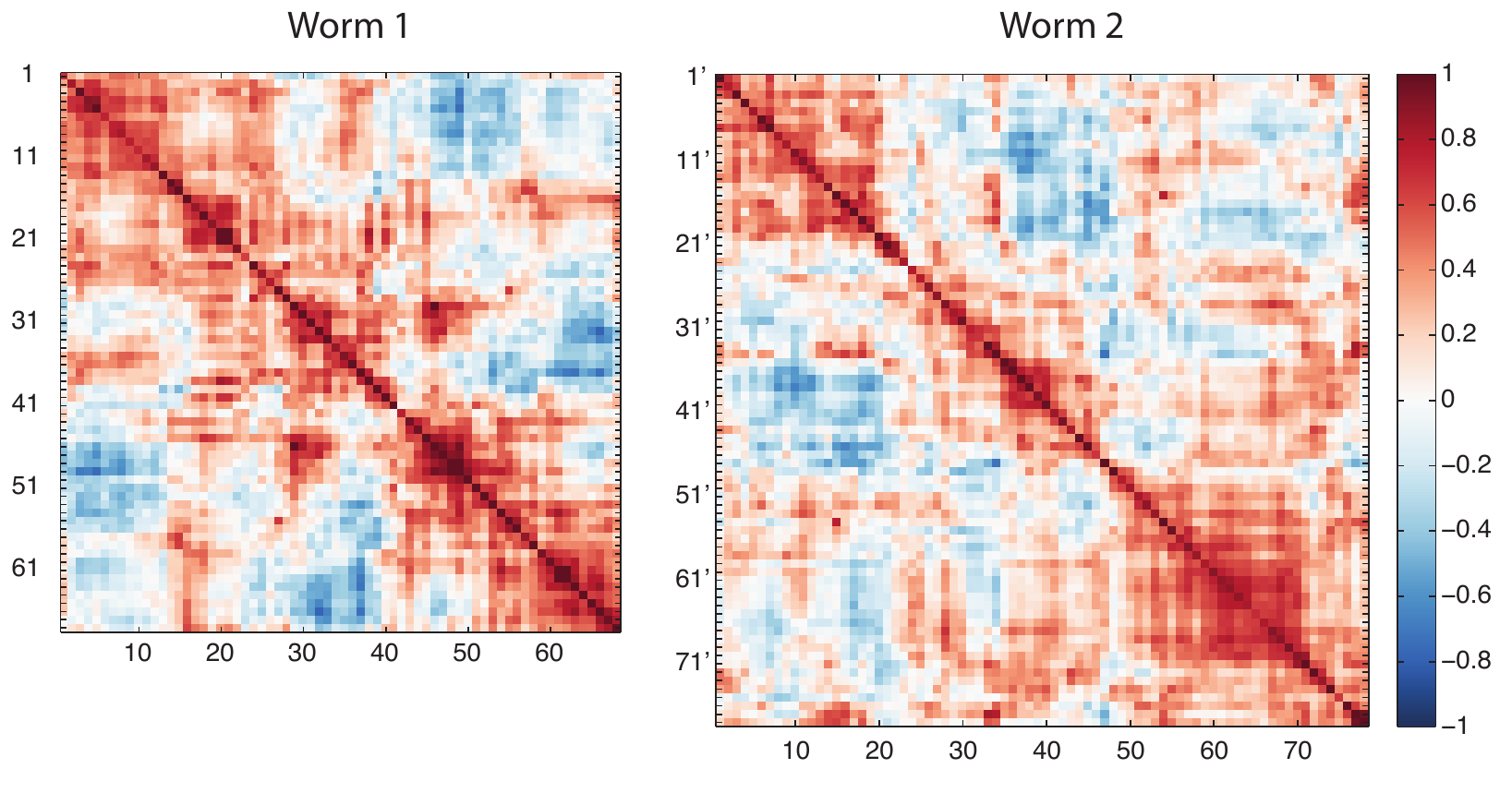}
\end{center}
\textbf{\refstepcounter{Sfig}\label{fig:corrMatrix} Supplementary Figure S\arabic{Sfig}. }{| Correlations between calcium activity of neurons from two worms is shown. Data corresponds to neural activity shown in Figure \ref{fig:heatmap}. Correlation values are heirarchically clustered using a Euclidean distance metric so that neurons with similar activity are organized together.}
\end{figure}

\begin{figure}
\begin{center}
\includegraphics[width=6in]{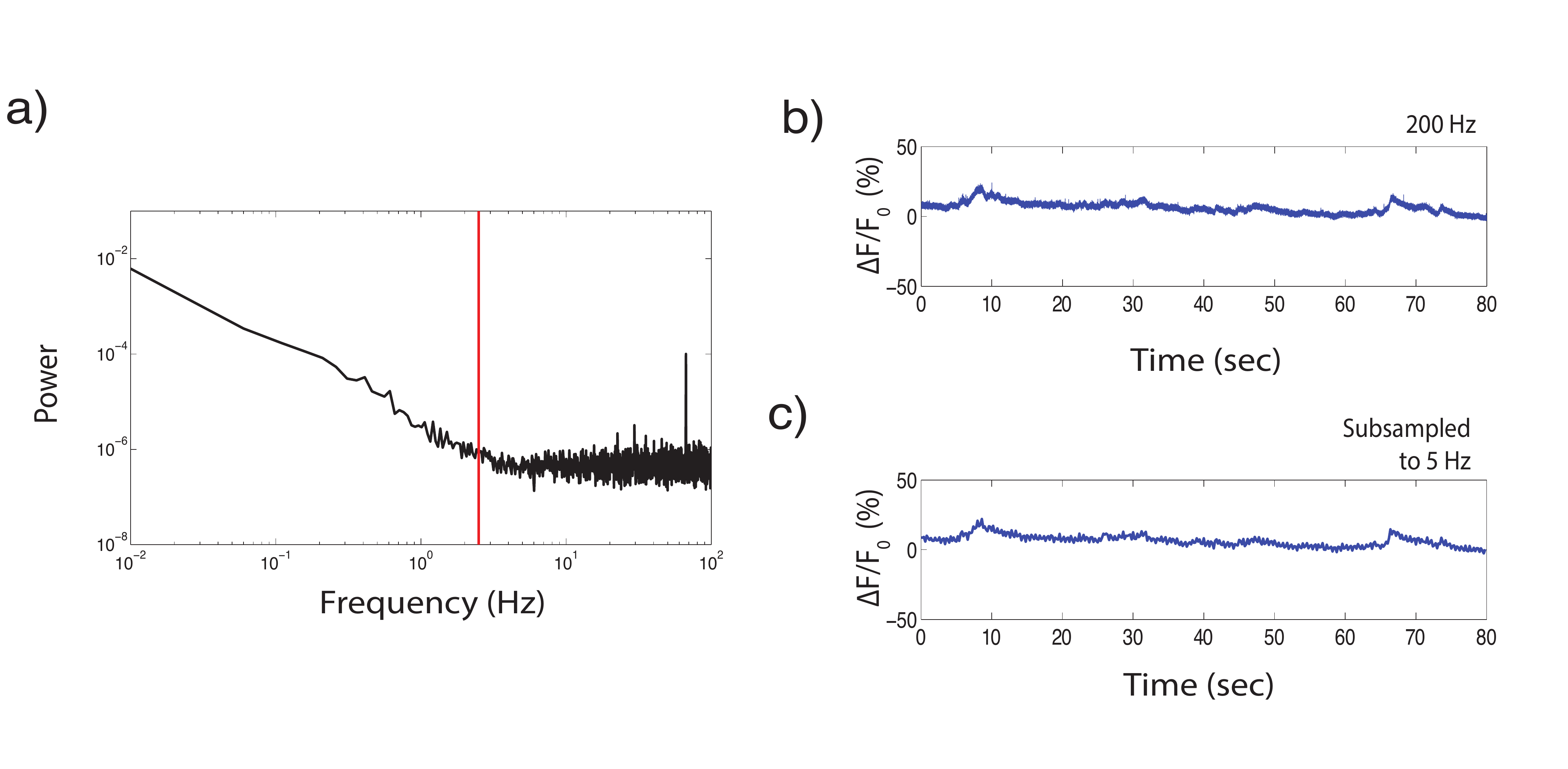}
\end{center}
\textbf{\refstepcounter{Sfig}\label{fig:PSD} Supplementary Figure S\arabic{Sfig}. }{| The system captures calcium dynamics. A single plane of an immobilized and awake worm was imaged at 200 fps. Calcium dynamics from 24 neurons in the imaging plane were recorded. \textbf{a)} Averaged power spectra of GCaMP6s fluorescent intensity from the 24 neurons are show. Red line indicates the Nyquist frequency of the whole brain imaging system (2.5 Hz, half the sampling rate). 86\% of the power is located below the Nyquist limit.  \textbf{b)} Calcium trace from a single neuron is shown recorded at 200 Hz \textbf{c)} The same trace is subsampled in software to 5 Hz to match the rate used in freely moving experiments.}
\end{figure}

\begin{figure}[htb]
\begin{center}
\includegraphics[width=5in]{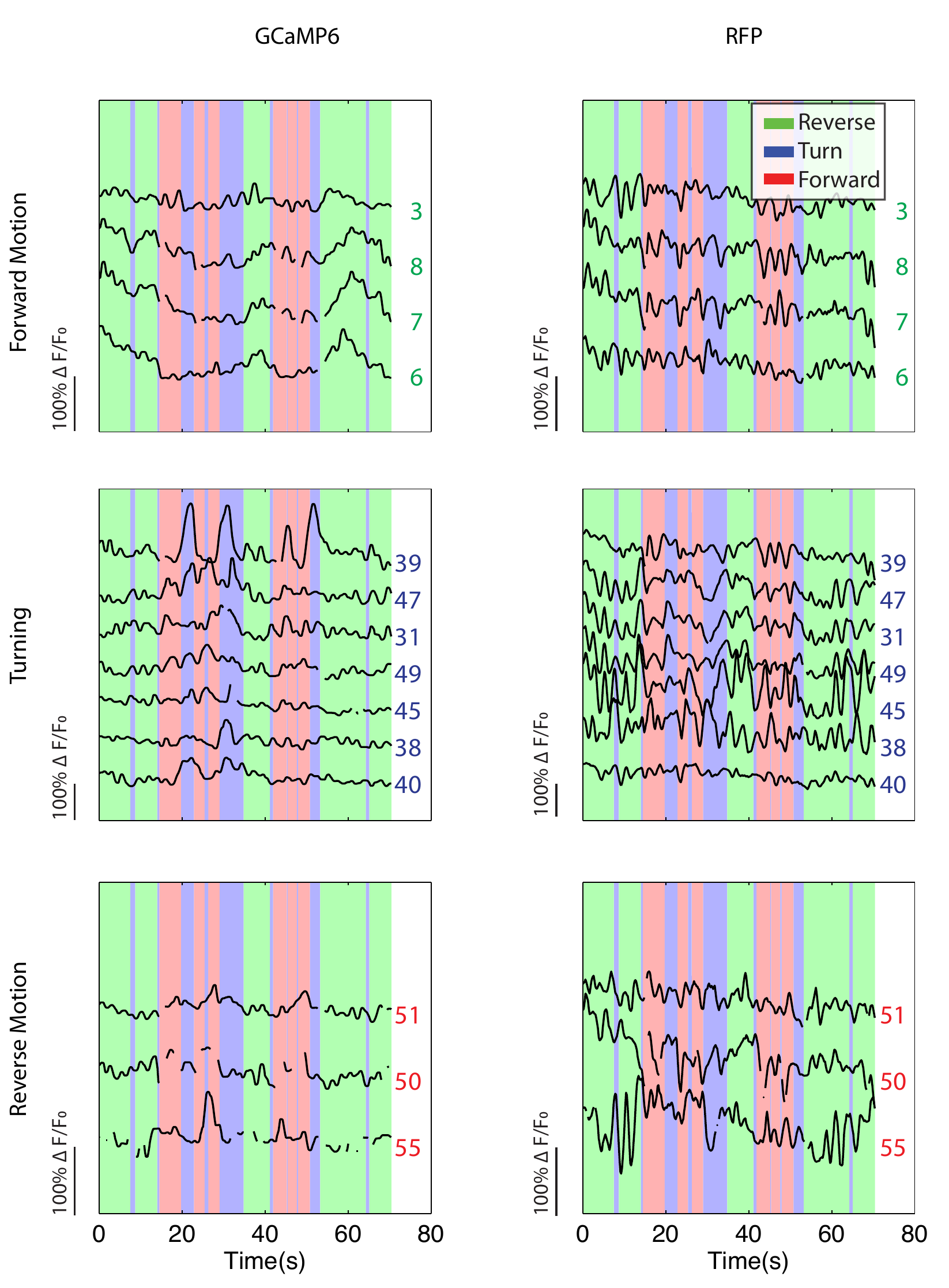}
\end{center}
\textbf{\refstepcounter{Sfig}\label{fig:rfpcontrol1} Supplementary Figure S\arabic{Sfig}. }{|  GCaMP6s and RFP fluorescence is shown for neurons that correlate with behavior in Worm 1. This data corresponds to the traces shown in Figure \ref{fig:behaviortraces}. The average correlation coefficients of traces shown here were  $r=0.45$ for GCaMP6s and $r=0.08$ for  RFP.}
\end{figure}

\begin{figure}[htb]
\begin{center}
\includegraphics[width=5in]{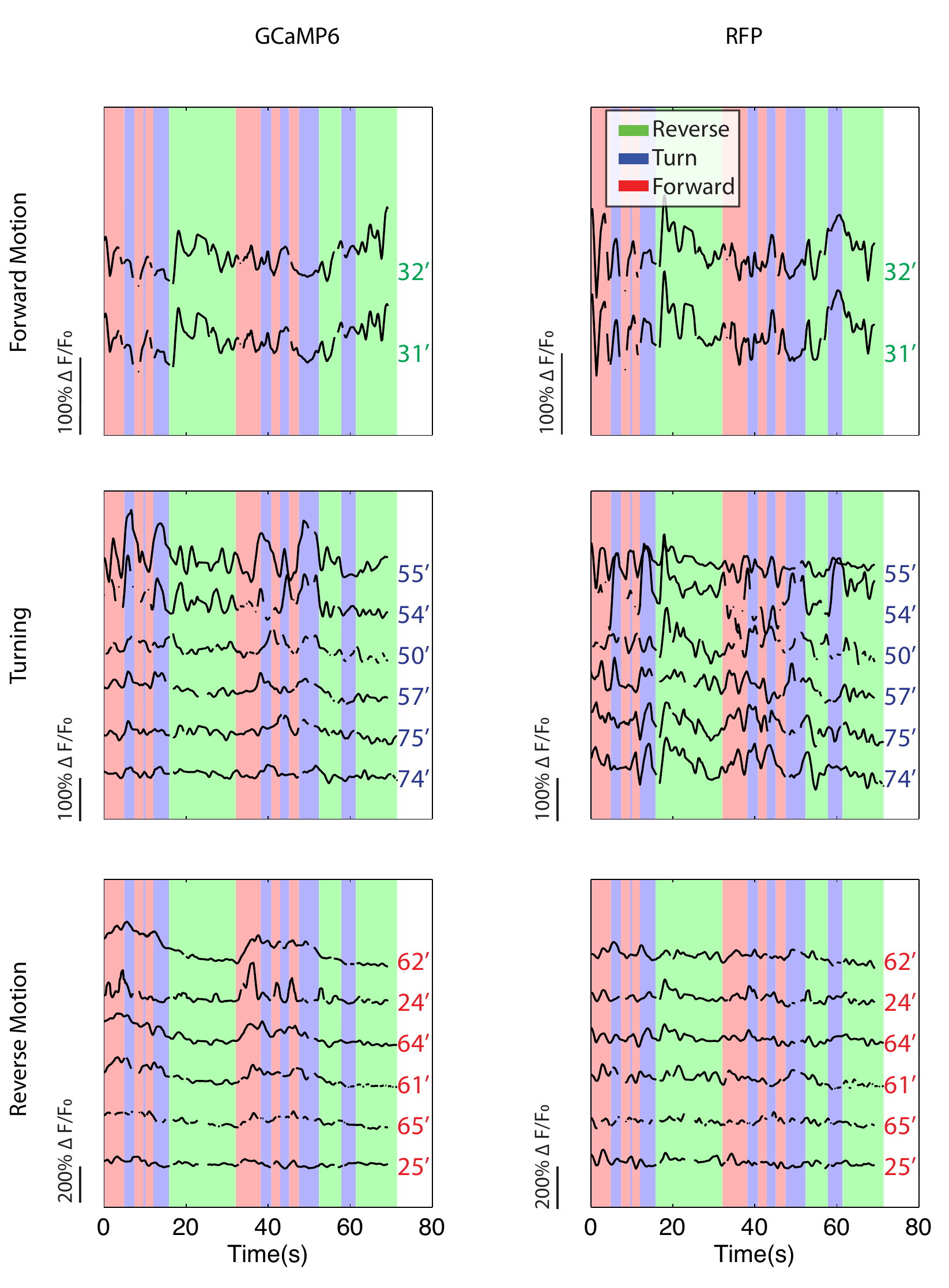}
\end{center}
\textbf{\refstepcounter{Sfig}\label{fig:rfpcontrol2} Supplementary Figure S\arabic{Sfig}. }{|  GCaMP6s and RFP fluorescence is shown for neurons that correlate with behavior in Worm 2. This data corresponds to the traces shown in Figure \ref{fig:behaviortraces}.  The average correlation coefficients of traces shown here were $r= 0.55$ for GCaMP6s and $r=0.19$ for RFP.} 
\end{figure}

\begin{figure}[htb]
\begin{center}
\includegraphics[width=3in]{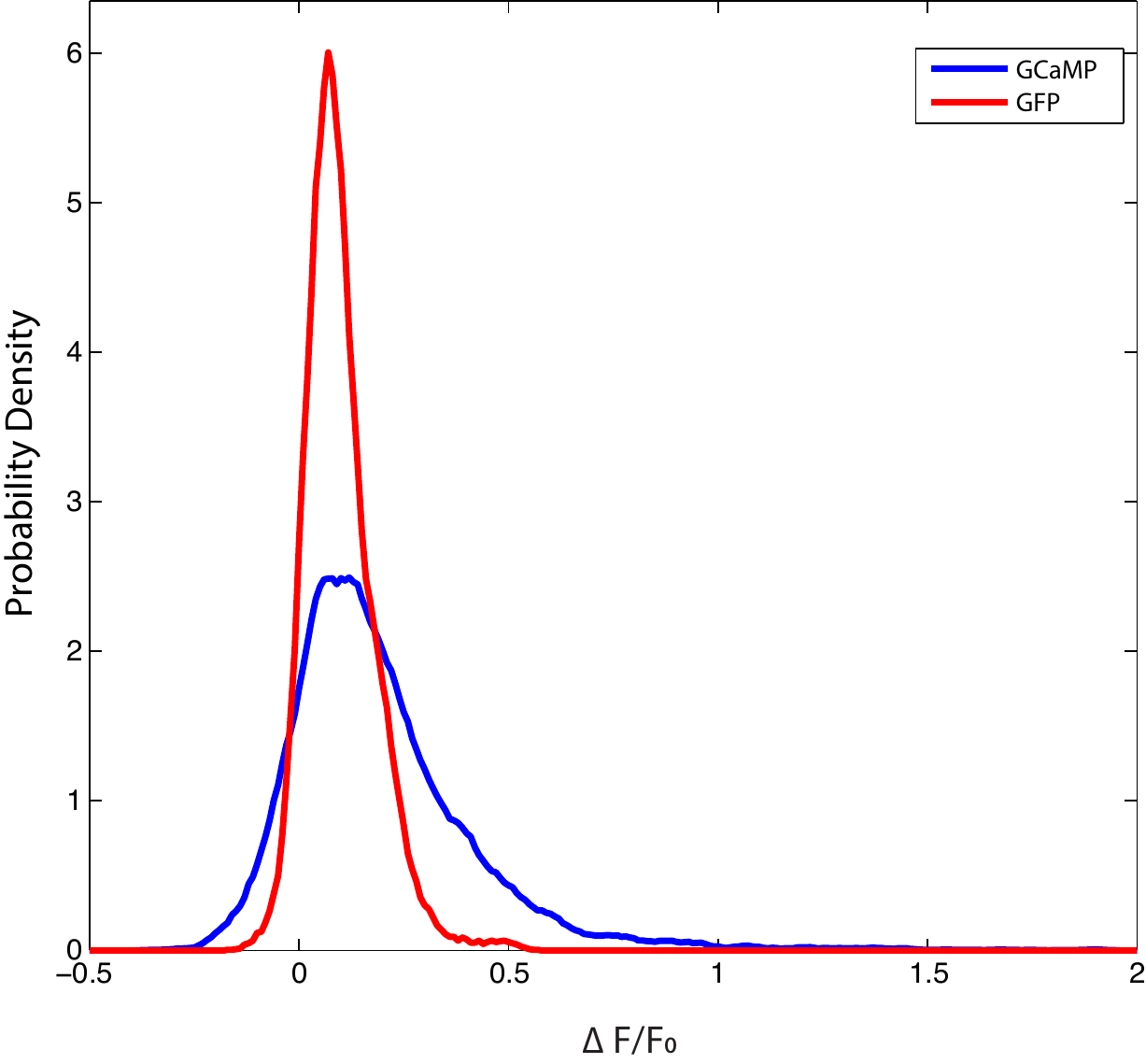}
\end{center}
\textbf{\refstepcounter{Sfig}\label{fig:GFPhist} Supplementary Figure S\arabic{Sfig}. }{| The  distribution of fluorescent intensity values,  $\Delta F / F_0$, of all neurons recorded during free behavior is shown for a GCaMP6s worm and a control GFP worm. The GCaMP6s had larger time-varying fluctuations in fluorescent intensity $\sigma_{\textrm{GCaMP}}=0.20$ than that of the  GFP worm $\sigma_{\textrm{GFP}}=0.08$. This is consistent with GCaMP6s' role as a calcium indicator and suggests that the time varying fluorescence we observe is not merely due to motion artifact.  The data shown for the GCaMP6s worm is the same as that in Figure \ref{fig:heatmap} (78 neurons recorded for 70 seconds). The control GFP worm  was recorded under similar conditions and exhibited qualitatively similar behaviors (43 neurons recorded for 23 seconds.)}
\end{figure}

A $ 8 x 8 x 8\mu m$ volume is cropped around each position and a threshold and area filter are used to detect the pixels of interest corresponding to each neuron's nucleus. The average  GCaMP6 intensity was calculated over the pixels to give the fluorsence intensity $F$. The neuron activity is expressed as the fractional change $\Delta F / F_0$, where the lower 20th percentile value is used as $F_0$. In certain frames, a combination of worm motion and tracking latency caused neurons to leave the field of view. These data were omitted from the analysis and appear as white gaps in plots of neural traces. Neurons were sorted by hierarchical clustering of the correlation matrices (Figure  S\ref{fig:corrMatrix}) using a Euclidean distance metric.

\subsection{Analyzing behavior}
Position, velocity and postural information were recorded during imaging sessions. The location of the motorized stage was recorded at 20 Hz which, when combined with the imaging, provides the worm's position and velocity within the arena. The worm's centerline also provided important behavioral information. The centerline of the worm was found in each low-magnification behavior image (61 fps) using a semi-automated active contour model, similar to \cite{Gong2013}, with slight modifications, whereby each contour consisted of 100 segments.  The first and last point of the contour were manually determined and fixed for each frame.  Additionally, a repulsive force between the first and last third of the contour was excluded when the two were in close proximity, so as to handle situations where the worm touched or crossed itself.  The initial contour of the first frame was determined manually; each consecutive frame used the final contour of the previous frame as its initial contour.  In instances where the algorithm failed to find the centerline of the worm, the parameters of the active contour model were adjusted until the centerline was found.

The speed of the worm was measured using a combination of stage position and the position of the center of mass of the worm centerline in the low magnification behavior image. The change in centerline curvature was used to determine whether the worms moved forward or backwards. We fit the centerline curvature of the worm to a sine wave and scored the velocity as negative when the wave propagated towards the head as opposed to the tail. 

The worm center-of-mass velocities were separated into three groups by k-means clustering to produce the behavior ethogram.  The clustering separated forward, backward, and turning states of the worm.  It was noted that the worms in this study only paused when executing deep bends and turns. Turning bouts shorter than 0.75 seconds were filtered from the ethogram.

\subsection{Correlating behavior with neural signals}

Pearson correlation coefficients, $r$, were calculated between each neural signal and each of the three behaviors. The $r$ values were compared with a noise model in order to determine significance. Each neural signal was shuffled 1500 times and a noise distribution of $r$ values was generated by comparing the shuffled data to the behavior. Shuffling was performed by separating the original trace into segments and randomizing their order. The size of the segments was equal to the correlation time of the original trace and was on the order 5s. Correlations with behavior were only  considered to be  significant if the $r$ value was greater than the largest of either $r=0.3$ or the 95th percentile of the noise distribution.  No neurons were found to be significantly correlated with more than one behavioral state.

\bibliographystyle{plain}
\bibliography{wholebrainBib}

\newpage
\section{Supplementary Material}

\newcounter{Sfig}

\begin{table}[htbp]
  \centering
    \begin{tabular}{ccccc}
    \toprule
    \textbf{Neuron type} & \textbf{Behavior} &  \textbf{No. in FOV} & \textbf{Evidence} & \textbf{Reference} \\
    \toprule
    AVB   & Forward & 2     & Calcium Imaging & \cite{Kawano2011} \\
    VB-type  & Forward & 1     & Calcium Imaging & \cite{Kawano2011, Piggott2011} \\
    \midrule
          & \textit{Total in FOV} & 3     &       &  \\
          &       &       &       &  \\
          \midrule
    AVA   & Reverse & 2     & Calcium imaging & \cite{BenArous2010,Kawano2011,Piggott2011,Faumont2011,Shipley2014} \\
    AVE   & Reverse & 2     & Calcium imaging & \cite{Kawano2011} \\
    AIB   & Reverse & 2     & Calcium imaging & \cite{Piggott2011} \\
    VA-type & Reverse & 1     & Calcium imaging & \cite{Faumont2011,Kawano2011} \\
    \midrule
          & \textit{Total in FOV} & 7     &       &  \\
          &       &       &       &  \\
          \midrule
    RIM   & Mixed Forward \\& Reverse & 2     & Calcium Imaging & \cite{Piggott2011} \\
    \midrule
          & \textit{Total in FOV} & 2     &       &  \\
          &       &       &       &  \\
          &       &       &       &  \\
          \midrule
    DD-type & Turning & 1     & Laser ablation & \cite{Donnelly2013} \\
    RIB   & Turning & 2     & Laser ablation & \cite{Gray2005} \\
    SMD   & Turning & 4     & Laser ablation & \cite{Gray2005} \\
    RIV   & Turning & 2     & Laser ablation & \cite{Gray2005} \\
    \midrule
          & \textit{Total in FOV} & 9     &       &  \\
    \bottomrule
    \end{tabular}%
  \label{tab:addlabel}%
    \caption{Summary of evidence in literature implicating neurons in forward, reverse and turning behavior. The neuron type is listed, as well as the behavior and the number of neurons whose location should be within our imaging field of view (FOV). For example, there are many VA motor neurons, but only one lies within our field of view. For forward and reverse behavior only evidence from calcium imaging studies are shown. For turning behavior the only calcium imaging studies in the literature focus on sensory neurons \cite{Larsch2013,Kato2014} which we are omitting here, so laser ablation studies were included. \label{table:NeuronEvidence}  }
 \end{table}

\begin{figure}[htb]
\begin{center}
\includegraphics[width=5.5in]{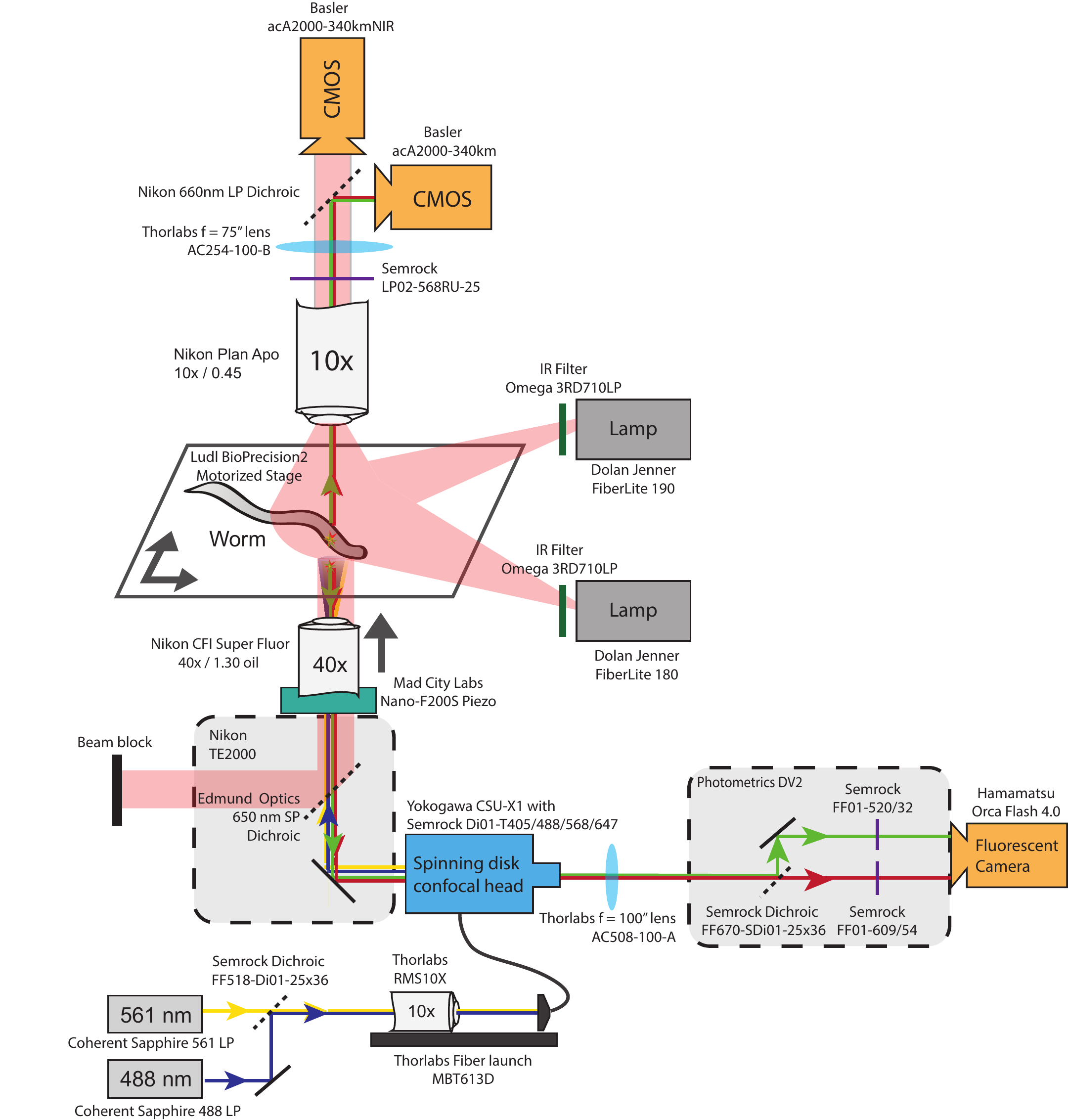}
\end{center}
\textbf{\refstepcounter{Sfig}\label{fig:fullOptics} Supplementary Figure S\arabic{Sfig}.}{ A schematic of the optical elements used.}
\end{figure}

\clearpage
\newpage
\textbf{Supplementary Movie S1.} Raw video recording of whole brain imaging of a freely moving worm. 200fps are shown here in real-time.  The animal expresses pan-neuronal nuclear localized RFP and pan-neuronal GCaMP6s. The worm crawls on an agarose pad and its position is tracked to keep its brain in the field of view. A piezoelectric stage moves the imaging plane in a triangular wave with a frequency of 2.5Hz, allowing us to image five volumes per second. As the objective scans through the worm's volume it goes from dark (outside the worm) to bright (inside the worm) to dark again (outside the worm), giving the appearance of 5 Hz flash.

Video available at: \url{http://leiferlab.princeton.edu/outsideAssets/wholebrain/Smovie1.mp4}

\textbf{Supplementary Movie S2.}  Simultaneous recording  of neural anatomy, calcium dynamics and  behavior of a freely moving worm expressing pan-neuronal nuclear localized RFP and pan-neuronal GCaMP6s.  Top left shows raw recording of high-magnification RFP image. Bottom left shows corresponding GCaMP6s. Both high-magnification images are computationally warped using 68 neuron locations as fiducial reference points in order to correct for motion and deformation of the worm's brain during behavior. Images are recorded as stacks in $z$, as in Supplementary Movie 1, but only a single image plane per stack is shown here. Video playback is real-time, but  the video  has been subsampled to 5 fps so as to show only a singe image plane in $z$. Fluorescent images are false colored. Top right shows  low magnification dark-field images of the worm, and the extracted centerline. The bottom right shows the position of the worm in the arena millimeters. The color of the circle corresponds with the classification of the behavior: green indicates that the worm is moving forward, red indicates it is moving backwards, and blue corresponds to deep turns made by the worm. All video streams have been subsampled from their original frame rates to 5 fps. 

Video available at: \url{http://leiferlab.princeton.edu/outsideAssets/wholebrain/Smovie2.mp4}

\end{document}